\RequirePackage{fix-cm}

\documentclass[]{interact}

\usepackage{epstopdf}
\usepackage[caption=false]{subfig}
\usepackage{subcaption}

\usepackage[natbibapa,nodoi]{apacite}
\theoremstyle{plain}

\theoremstyle{definition}

\theoremstyle{remark}

\usepackage[ruled,vlined]{algorithm2e}
\usepackage{multicol}
\usepackage{hyperref}
\usepackage[noabbrev,capitalize]{cleveref}
\creflabelformat{equation}{#2#1#3}

\begin{document}

\title{XFlowMap: Cross-Scale Generalization and Mapping of Massive Origin-Destination Data}

\author{
\name{Diansheng Guo$^*$\thanks{$^*$ Corresponding author. Email: diansheng.guo@polyu.edu.hk} and Hai Jin}
\affil{Department of Land Surveying and Geo-Informatics, The Hong Kong Polytechnic University, Kowloon, Hong Kong}
} 

\maketitle

\begin{abstract}
Mapping large origin-destination (OD) datasets remains challenging because flow maps become cluttered, meaningful patterns occur at multiple spatial scales, and existing flow-mapping approaches frequently rely on predefined aggregation units or manual generalization. This paper presents \textit{XFlowMap}, a framework for the cross-scale generalization and mapping of massive OD data. Specifically, the framework integrates cross-scale flow pattern (cluster) detection, automated flow map generalization, and a new cartographic representation for analyzing and visualizing complex origin-destination flow structures. The approach detects salient flow patterns at their appropriate origin and destination scales, extracts high-level structures, and generates a new flow map representation that supports holistic interpretation of complex origin-destination flow patterns. A scan-statistic-based procedure is developed to evaluate and generalize cross-scale flow clusters. The detected clusters are then visualized using a novel flow symbol that integrates location, direction, strength, and OD scales in a single representation. The framework supports both area-based and point-based OD data, is robust to sparse and noisy datasets, and enables comparative mapping of stratified flow data. Experiments with synthetic data and U.S. migration data demonstrate that the method effectively extracts meaningful cross-scale flow patterns and produces clear, information-rich flow maps for large mobility datasets, supporting both static presentation and interactive exploration.
\end{abstract}


\section{Introduction}

Flow maps have long been used to visualize geographic movements such as commodity flows, human migration, and traffic flows. Large mobility datasets are now available from many sources such as census surveys, mobile phones \citep{gaoDiscoveringSpatialInteraction2013}, geo-tagged social media \citep{guoDetectingNonpersonalSpam2014,liuUncoveringPatternsInterUrban2014}, taxi trips \citep{guoDiscoveringSpatialPatterns2012,liuUnderstandingIntraurbanTrip2012}, and simulation model outputs \citep{eubankModellingDiseaseOutbreaks2004,guoVisualAnalyticsSpatial2007}. However, effectively analyzing and mapping such data remains challenging, particularly when synthesizing massive mobility datasets into meaningful flow maps in an automated way. It requires effective methods for flow pattern extraction, objective evaluation of candidate patterns, and flow map generalization. Most professional flow maps produced today still rely heavily on human interaction and subjective decisions to aggregate data and generalize patterns, often at arbitrarily chosen spatial scales.

There are three major challenges for flow mapping. \emph{First}, a flow map can quickly become cluttered and illegible when many flows are plotted in a limited map space---a problem commonly referred to as visual clutter. Effective generalization approaches are therefore needed to synthesize massive flow data into high-level flow patterns automatically. \emph{Second}, spatial flow patterns may occur at different scales and across scales. For example, one migration trend may occur between large regions (such as flows from the Midwest to the Southwest in the U.S.), while another may occur between relatively small areas (e.g., from New York City to Miami). Existing methods usually analyze or aggregate flows using predefined spatial units, which may not correspond to the inherent patterns in the data. \emph{Third}, flows across different spatial extents are difficult to compare because the sizes of origin and destination regions may vary greatly. While a number of methods have been proposed in recent years (reviewed in \cref{sec:background}), relatively little research has focused on automated, multi-scale and cross-scale mapping of large mobility datasets.

Addressing these challenges requires methods that can extract meaningful flow patterns at their appropriate spatial scales and present them in a holistic and interpretable cartographic representation. This paper presents a framework, \textit{XFlowMap}, for automated cross-scale flow mapping and generalization of OD data. The contributions of this paper are threefold. First, we introduce a method for extracting salient origin-destination flow patterns from massive OD datasets to support automated flow map generalization. Second, it determines appropriate spatial scales for both origin and destination regions through a cross-scale search procedure, enabling the detection of flow patterns that span different spatial extents. Third, we present a new flow map representation that integrates origin location, destination location, scale, direction, and strength of flows in a single visual form, allowing complex OD patterns to be mapped and interpreted holistically. The method is robust to sparse and noisy data and can also be used to compare flow patterns across stratified datasets, such as flows of different population groups or time periods. We evaluate the method through two case studies: one using a synthetic point-based dataset and another using a county-to-county migration dataset for the United States.

\section{Background}
\label{sec:background}

Traditional flow maps were primarily designed to visualize movements among a relatively small set of locations \citep{toblerSpatialInteractionPatterns1976,toblerModelGeographicalMovement1981,poonCosmopolitanizationTradeRegions1997,youngNewSpaceTimeComputer2002,mitchellIdentifyingFunctionalRegions2010}. As spatial mobility datasets have grown rapidly in size and complexity, there is an increasing need to generalize large origin-destination (OD) datasets and render meaningful flow maps in an automated way \citep{guoVisualAnalyticsSpatial2007,guoFlowMappingMultivariate2009,guoDiscoveringSpatialPatterns2012,guoOriginDestinationFlowData2014,andrienkoVisualAnalyticsMobility2017,wang2014pattern}. Flow mapping for large mobility datasets presents several major challenges. First, a flow map can quickly become cluttered and illegible when many flows are plotted in a limited map space—a problem commonly referred to as visual clutter. Second, flows often occur among geographic units (e.g., counties) that differ dramatically in size (e.g., population), making flows or clusters of flows difficult to compare directly. Third, flow patterns may exist at multiple spatial scales, often extending beyond the spatial units provided in the data. Finally, there remains the broader challenge of automated flow map generalization—how to extract meaningful structures from large mobility datasets and present a holistic view of complex flow patterns at appropriate spatial scales.

A number of visualization approaches have been developed to address the visual cluttering problem. One early approach is the vector flow surface proposed by \citet{toblerExperimentsMigrationMapping1987}, which decomposes long-range flows into sequences of short flows between adjacent locations. This representation reveals general flow directions but loses the explicit origin-destination relationships of individual flows. Another class of approaches re-route flows or bundle edges to improve visual clarity in flow maps \citep{phanFlowMapLayout2005,cuiGeometryBasedEdgeClustering2008,holtenForceDirectedEdgeBundling2009,buchinFlowMapLayout2011,zengRouteAwareEdgeBundling2019}. These methods can reduce visual clutter by merging nearby flows or altering their routes, but they may weaken the visual cue of direct connections between origins and destinations. Other alternative visualization techniques for presenting flow data include matrix views (often coordinated with map views) \citep{guoVisualAnalyticsSpatial2007} and abstract map matrices such as the \textit{Map\textsuperscript{2}} matrix \citep{guo2006VISSTAMP}, which was later adapted by OD-Map \citep{woodVisualisationOriginsDestinations2010}.

Another common strategy for reducing visual clutter in flow mapping is location aggregation. This may involve using higher-level administrative units, spatial aggregation, or graph partitioning \citep{guoFlowMappingMultivariate2009}. Flow maps are then constructed based on flows among these aggregated regions. More recent work has proposed methods that identify and map clusters of flows \citep{zhuMappingLargeSpatial2014,taoFlowHDBSCANHierarchicalDensityBased2017,yaoStepwiseSpatioTemporalFlow2018,songDetectingArbitrarilyShaped2019,liuSNN_flowSharedNearestneighborbased2022}. While these approaches can extract higher-level structures from large OD datasets, different aggregation schemes may lead to different patterns, a problem commonly known as the modifiable areal unit problem (MAUP) \citep{openshawModifiableArealUnit1984}. In addition, flows among clusters of different sizes are often difficult to compare in flow maps due to the lack of normalization mechanisms.

To address both the cluttering and normalization challenges, \citet{guoOriginDestinationFlowData2014} proposed a flow density estimation approach. This method removes the size effect among spatial units using kernel-based flow density estimation. A flow density is calculated for each pair of origin and destination using the same population bandwidth (for example, one million population at both the origin and the destination). Major flows with high flow density are then extracted, and overlapping flows are filtered out to enable effective mapping of high-level flow patterns without bundling or altering flow paths. By adjusting the population bandwidth, the method can reveal flow patterns at different spatial scales. However, the method is best suited for detecting flow patterns at a chosen scale determined by the bandwidth and does not directly identify patterns across scales, for example, flows between regions of substantially different population sizes.

This flow density estimation approach differs from other location-based density representations, such as raster-based density maps that count the number of flow lines passing through each grid cell \citep{raeSpatialInteractionData2009} or smoothed location measures derived from flows \citep{scheepensCompositeDensityMaps2011,guoDiscoveringSpatialPatterns2012,scheepensInteractiveDensityMaps2012,koyluSmoothingLocationalMeasures2013}. In addition, several recent studies have proposed statistical approaches for analyzing spatial structures in flow data, including adaptations of spatial point-pattern statistics and flow interaction measures \citep{taoFlowCrossKfunction2019,kanRipleysKfunctionNetworkConstrained2022,shuLfunctionGeographicalFlows2021,yanSpatiotemporalFlowLfunction2023}.

Despite these advances, there remains a critical need for effective flow mapping of massive origin-destination data to enable a holistic understanding of patterns across different spatial scales. Relatively little research has addressed multi-scale and cross-scale flow mapping and generalization. Two major challenges arise when attempting to discover and map cross-scale flow patterns. First, flows among regions of different sizes are difficult to compare and generalize. Second, incorporating multiple spatial scales into the same map substantially increases the complexity of pattern representation and interpretation, as each candidate pattern involves an origin region, a destination region, a flow magnitude (or derived statistic), and potentially different spatial scales at both the origin and the destination.

One framework that can support the systematic search and statistical evaluation of spatial patterns is the scan statistic. Originally developed for detecting spatial or space-time clusters of events, scan statistics evaluate candidate regions by comparing the observed number of events with their expected values under a null model. A few recent studies have adapted spatial scan statistics to detect flow clusters in origin-destination data \citep{wang2014pattern,songDetectingArbitrarilyShaped2019}. In this research, we adapt the three-dimensional space-time permutation scan statistic \citep{kulldorffSpaceTimePermutation2005} for cross-scale flow map generalization. The original space-time scan statistic \citep{kulldorffProspectiveTimePeriodic2001,kulldorffSpaceTimePermutation2005} scans space-time events using a three-dimensional cylindrical window to detect clusters. The base of the cylinder is a circular window in geographic space, while the height represents a temporal window. By varying the cylinder's location, base radius, and height, the method searches for clusters across both space and time. Under the Poisson model, the strength of a candidate cluster is evaluated using a likelihood ratio statistic. For a candidate cylinder \(Z\), the likelihood ratio (LR) is defined as \citep{kulldorffProspectiveTimePeriodic2001}:
\begin{equation}
LR =
\begin{cases}
\left( \frac{n_z}{\mu_z} \right)^{n_z}
\left( \frac{N-n_z}{N-\mu_z} \right)^{N-n_z}, & n_z > \mu_z \\
1, & n_z \le \mu_z
\end{cases}
\end{equation}
where \(N\) is the total number of events in the dataset, \(n_z\) is the observed number of events in cylinder \(Z\), and \(\mu_z\) is the expected number of events under the null hypothesis. Higher values of \(LR\) indicate stronger evidence of clustering. The expected value \(\mu_z\) can be estimated under different null models depending on the application. In the following section, we extend this framework to analyze origin-destination flows by incorporating spatial scales at both the origin and the destination.

\section{XFlowMap: Cross-Scale Flow Pattern Generalization and Mapping}

\subsection{Problem Definition and Overview}

\textit{XFlowMap} is designed to generalize large origin-destination (OD) flow datasets and generate a new flow map representation that reveals salient cross-scale flow patterns in a holistic manner. Instead of mapping all individual flows directly, the method proceeds in three steps. First, it identifies statistically strong flow structures between origin and destination regions at appropriate spatial scales using an extension of spatial scan statistics. Second, it filters redundant patterns and derives a generalized representation of the overall flow structure across scales. Third, the selected patterns are rendered using a new flow map representation that synthesizes multiple dimensions of information for each flow cluster, including the locations of origin and destination, the spatial scales of the two regions, the strength of the flow pattern (represented by the logarithmic generalized likelihood ratio---\(\mathit{LGLR}\)), and flow direction. In addition, \textit{XFlowMap}  is designed to be computationally efficient so that it can process massive OD datasets while supporting interactive exploration.

Let \(T = \{f_{od}\}\) denote an origin-destination (OD) flow dataset defined over a study area represented by a finite set of locations \(S\). Let \(n=|T|\) be the total number of flows, and \(m=|S|\) be the number of unique locations. The locations may correspond to point locations or spatial units (represented by their centroids). Each flow \(f_{od} \in T\) represents a directed movement from an origin \(o \in S\) to a destination \(d \in S\) with flow volume \(f_{od}\). The total flow volume in the dataset is \(F = \sum_{o \in S}\sum_{d \in S} f_{od}\). Each location \(s_i \in S\) may have associated attributes such as population \(p_i\), total inflow \(F_{*i} = \sum_{o \in S} f_{oi}\), and total outflow \(F_{i*} = \sum_{d \in S} f_{id}\), which can be used to evaluate candidate flow patterns.

\begin{figure}[htbp]
    \centering
    \includegraphics[width=0.5\textwidth]{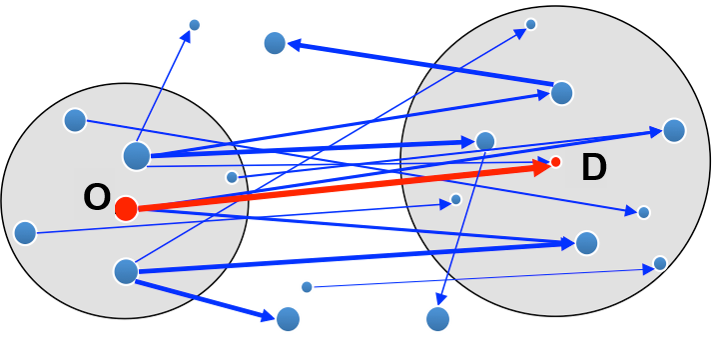}
    \caption{Illustration of a candidate flow pattern (cluster), which is defined by a circular origin area \(O\), a circular destination area \(D\), and all flows from \(O\) to \(D\).}
    \label{fig:clusterExample}
\end{figure}

A candidate flow pattern (or flow cluster) is defined by an origin neighborhood \(O \subset S\), a destination neighborhood \(D \subset S\), and the set of flows from \(O\) to \(D\) (see Figure \ref{fig:clusterExample}). In this research, both \(O\) and \(D\) are represented as circular neighborhoods centered on an origin location \(o\) and a destination location \(d\), respectively. In implementation, the origin neighborhood \(O\)(or \(D\)) is constructed as the set of the \(k\)-nearest neighbors of \(O\) (or \(D\)). By increasing \(k\) from \(1\) to \(\mathit{maxK}\), the method enumerates neighborhoods of increasing size (scale), where \(\mathit{maxK}\) is typically set to \(m/2\). A candidate pattern therefore has four essential components: origin location, destination location, origin scale, and destination scale. 

Circular neighborhoods are adopted because they maintain strong detection power for clusters---even for irregularly shaped clusters---while enabling efficient enumeration of neighborhoods across scales \citep{duczmalEvaluationSpatialScan2006}.

The overall \textit{XFlowMap} framework consists of three main stages:
\begin{enumerate}
\item \textbf{Cross-Scale Search and Statistical Evaluation.} Candidate flow patterns are generated by scanning origin and destination neighborhoods across multiple scales around observed flows. Each candidate pattern is evaluated using a statistical measure to quantify its strength.
\item \textbf{Flow Pattern Generalization.} Candidate flow clusters are ranked according to their statistical strength, and a subset of salient and non-redundant patterns is selected to represent the overall flow structure.
\item \textbf{Cartographic Design for Cross-Scale Flow Mapping.} The selected patterns are rendered using a cross-scale flow map representation that supports both static visualization and interactive exploration.
\end{enumerate}

\subsection{Cross-Scale Search and Statistical Evaluation of Flow Clusters}
\label{subsec:scan}

To discover salient cross-scale flow patterns, \textit{XFlowMap} searches candidate flow clusters by exploring combinations of origin and destination neighborhoods across spatial scales. For each observed flow \(f_{od} \in T\), systematically varying the sizes of its origin and destination neighborhoods \(O\) and \(D\) generates different candidate flow clusters.

In principle, the neighborhood size for \(O\) or \(D\) could grow to include all locations in the dataset. However, such large neighborhoods are both unnecessary and undesirable for pattern detection. First, once a neighborhood covers a large portion of the dataset, the observed flow \(F_{OD}\) will approach its expected value under the null model, making it unlikely to produce statistically significant clusters. Second, overly large neighborhoods excessively aggregate data and reduce the spatial resolution of detected flow patterns. Therefore, it is sufficient to limit the maximum neighborhood size during the search.

In practice, we impose an upper bound on neighborhood size using either the number of locations or the total flow involved in a cluster. Specifically, the maximum neighborhood size can be set as \(\mathit{maxK} = |S|/5\) when measured by the number of spatial units, or \(\mathit{maxSize} = F/5\) when measured by the total flow volume involved in the cluster. These limits substantially reduce the search space while maintaining sufficient spatial resolution for detecting meaningful flow patterns.

\begin{figure}[htbp]
    \centering
    \includegraphics[width=0.8\textwidth]{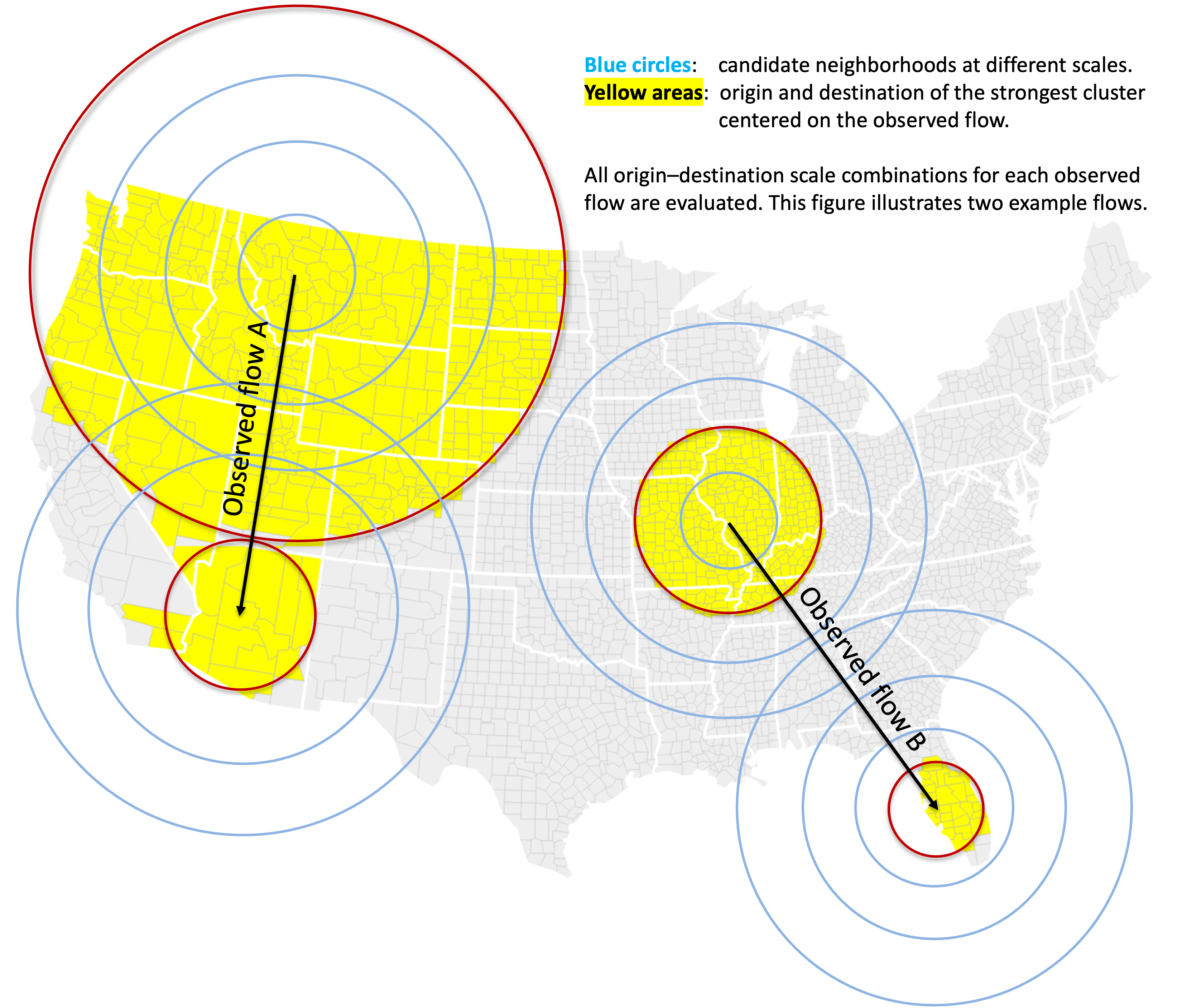}
    \caption{Cross-scale search of candidate flow clusters. For each observed flow (black arrows), concentric neighborhoods centered at the origin and destination represent candidate spatial scales (blue circles). Flow clusters are evaluated for all combinations of origin and destination scales. The highlighted regions (yellow) indicate the spatial units included in the strongest detected clusters for two example flows, with their scale circles shown in red.}
    \label{fig:XScanIllustration}
\end{figure}

With this constraint, each observed flow \(f_{od}\) can generate candidate flow clusters by varying the scales of its origin and destination neighborhoods. If \(\mathit{maxK}\) is used, each location can produce up to \(\mathit{maxK}\) neighborhoods, and an observed flow may therefore generate up to \(\mathit{maxK}^2\) candidate flow clusters corresponding to different combinations of origin and destination scales (\cref{fig:XScanIllustration}). The search therefore explores a four-dimensional space consisting of origin location, destination location, origin scale, and destination scale.

For each candidate flow cluster, we evaluate its strength by comparing the observed flow between the two regions with its expected value under a null model. Let \(F_{OD} = \sum_{o \in O}\sum_{d \in D} f_{od}\) denote the total observed flow from \(O\) to \(D\). Let \(F_{O*}\) be the total outflow from \(O\), \(F_{*D}\) the total inflow to \(D\), and \(F\) the total flow in the dataset.

Following the space-time scan statistic framework \citep{kulldorffProspectiveTimePeriodic2001,kulldorffSpaceTimePermutation2005}, we adopt a null model in which the marginal totals of inflow and outflow are fixed and origins and destinations are assumed to be independent. Under this model, the expected flow from \(O\) to \(D\) is
\begin{equation} \label{eq:flow_null_model}
    {\widehat{F}}_{OD} = (\frac{F_{O*}}{F})(\frac{F_{*D}}{F})F = \frac{F_{O*}F_{*D}}{F}
\end{equation}

Under this interaction-based model, \(F_{OD}\) follows a hypergeometric distribution with mean \({\widehat{F}}_{OD}\) \citep{riceMathematicalStatisticsData2007}. When \(F_{O*}\) and \(F_{*D}\) are small relative to the total flow \(F\), the distribution can be approximated by a Poisson distribution with mean \({\widehat{F}}_{OD}\) \citep{evansStatisticalDistributions2000,kulldorffSpaceTimePermutation2005}.

To measure the strength of a candidate pattern, we compute a generalized likelihood ratio (GLR) statistic similar to that used in scan statistics:
\begin{equation} \label{eq:glr}
{GLR}_{OD} =
\left[
\frac{F_{OD}}{{\widehat{F}}_{OD}}
\right]^{F_{OD}}
\left[
\frac{F - F_{OD}}{F - {\widehat{F}}_{OD}}
\right]^{F - F_{OD}}, \quad \text{if } F_{OD} > {\widehat{F}}_{OD}
\end{equation}

Because the focus of this study is on clusters of elevated flows, we set \({GLR}_{OD} = 1\) when \(F_{OD} \le {\widehat{F}}_{OD}\). For computational convenience, we use the logarithmic value of \(GLR\), denoted as \(\boldsymbol{LGLR}\), as the final test statistic. Larger \(\mathit{LGLR}\) values indicate stronger flow clusters regardless of the sizes of the origin and destination regions.

\cref{alg:scan} outlines the search algorithm, which evaluates candidate clusters around each observed flow and identifies the strongest cluster centered on that flow.

\vspace{1em}
\begin{algorithm}[H]
\caption{Flow Cluster Scan}
\label{alg:scan}
\KwIn{A dataset of flows $\{f_{od}\}$}
\KwOut{A set of clusters $C = \{C_v\}$}
\ForEach{flow $f_{od}$}{
    Let $\{O_i\}$ be all possible neighborhoods centered on origin $o$\;
    Let $\{D_k\}$ be all possible neighborhoods centered on destination $d$\;
    $C_{od} = \langle O, D, l=0 \rangle$ holds the cluster with the largest $\mathit{LGLR}$ value\;
    \ForEach{neighborhood $O_i$}{
        \ForEach{neighborhood $D_k$}{
            \If{$O_i$ and $D_k$ do not spatially overlap}{
                Calculate the $\mathit{LGLR}$ value from $O_i$ to $D_k$\;
                \If{$\mathit{LGLR} > C_{od}.l$}{
                    $C_{od} = \langle O_i, D_k, l = \mathit{LGLR} \rangle$\;
                }
            }
        }
    }
    \If{$C_{od}.l > 0$}{
        add $C_{od}$ to $C$\;
    }
}
\end{algorithm}
\vspace{1em}

The time complexity of \cref{alg:scan} is \(O(nm^{2})\), where \(n=|T|\) is the number of flows (not the flow volume) and \(m=|S|\) is the number of spatial units. For the U.S. migration data for age group 65-69 (see \cref{sec:migration}), which has \(F\)=916856 migrants, \(n\)=70131 county-to-county flows, and \(m\)=3075 counties, the running time is around one minute with a single CPU core (2.6 GHz Intel Core i7) and \(maxSize = 200000\) (in terms of the total flow involved in the neighborhood). 

To further speed up for empirical uses, it is straightforward to parallelize the scanning algorithm to employ multi-score processing or cloud-computing platforms since each evaluation is independent. For much larger datasets, such as billions of OD taxi trips and GPS points in the Manhattan (New York City) collected over multiple years, it is not necessary to scan every possible pair of GPS points or enumerate all possible neighborhoods by adding one GPS point at a time. Our strategy is to perform a preliminary k-means clustering to group massive GPS points into a set of small spatial clusters and then aggregate the initial data into a cluster-based flows, which substantially reduces the data size while preserves data patterns. For example, if we aggregate billions of GPS points in Manhattan (NYC) into 10000 k-means clusters, each cluster has a radius of about 50-100 meters, sufficient to represent the original data patterns. Then our approach can be applied to the aggregated data, where each location can still have thousands of different neighborhoods (scales).

\subsection{Flow Pattern Filtering and Generalization}
\label{subsec:generalization}

The cross-scale search described above generates a set of flow clusters, consisting of the strongest cluster identified for each observed flow. These clusters are naturally ranked by their strength, measured by the test statistic \(\mathit{LGLR}\). Clusters with larger \(\mathit{LGLR}\) values represent stronger deviations from the null model and therefore more salient flow patterns. 

In practice, the number of detected clusters is typically far larger than what can be effectively displayed in a map. Many clusters may also convey redundant information because their origin and destination neighborhoods overlap substantially. To produce a concise and interpretable representation of the overall flow structure, we select a subset of representative clusters as a generalization of the overall flow patterns. \cref{alg:selection} presents this selection process.

Specifically, all flow clusters are first ranked in descending order of \(\mathit{LGLR}\). We then select clusters sequentially from the top of the list while enforcing a non-overlap constraint. A candidate cluster is selected if its origin and destination neighborhoods do not overlap with those of any previously selected cluster at both ends. Two clusters are therefore considered overlapping only when their origin neighborhoods intersect and their destination neighborhoods intersect simultaneously. This rule allows clusters to share either origins or destinations while preventing redundant clusters that represent essentially the same flow pattern.

The selection procedure continues until no additional clusters can be added without violating the non-overlap constraint. The resulting subset forms a generalized representation of the dominant flow patterns in the dataset. Because clusters are processed in order of decreasing \(\mathit{LGLR}\), the selected clusters correspond to the strongest and most representative flow structures discovered in the cross-scale search.  

\vspace{1em}
\begin{algorithm}[H]
\caption{Flow Cluster Selection}
\label{alg:selection}
\KwIn{A set of flow clusters $C = \{C_v\}$}
\KwOut{A set of non-overlapping significant flow clusters $\{C_s\}$}
$\{C_s\} = \varnothing$\;
Sort $\{C_v\}$ by descending $LGLR$\;
\ForEach{flow cluster $C_v$, starting from the most significant one}{
    $\mathit{overlapFlag} = \text{false}$\;
    \ForEach{previously selected flow cluster $C_s$ in $\{C_s\}$}{
        \If{($C_v$ and $C_s$ overlap at origin $\mathit{AND}$ at destination)}{
            $\mathit{overlapFlag} = \text{true}$\;
        }
    }
    \If{$\mathit{overlapFlag}$ is false}{
        Add $C_v$ to $\{C_s\}$\;
    }
}
\end{algorithm}
\vspace{1em}

The time complexity of \cref{alg:selection} is \(O(nc)\), where \(n=|F|\) is the number of input flow clusters (one for each observed flow) and \(c\) is the number of selected clusters (\(c \ll n\)). The computational cost of this step is negligible compared with the cross-scale search.

Although the ranking of clusters is determined by \(\mathit{LGLR}\), statistical significance testing can optionally be applied to evaluate the reliability of detected clusters. Following the scan statistic framework, a Monte Carlo procedure can be used to generate the empirical distribution of the maximum \(\mathit{LGLR}\) value under the null hypothesis. For the null model defined in \cref{eq:flow_null_model}, the original flow data are permuted using an unbiased Fisher--Yates shuffle (\cref{alg:permutation}). The permutation guarantees that the marginal totals of inflow and outflow for each location remain unchanged. Each permuted dataset is scanned to obtain the maximum \(\mathit{LGLR}\) value among all candidate clusters. Repeating this process generates an empirical distribution of the maximum test statistic under the null model.

\vspace{1em}
\begin{algorithm}[H]
\caption{Data Permutation}
\label{alg:permutation}
\KwIn{the original dataset of $n$ flows $\{f_i\}, i = 1 \dots n$}
\KwOut{a permutated dataset of $n$ ``new'' flows $\{f_i'\}$}
\Repeat{a desired number of times (e.g., 10 times)}{
    \For{$i \leftarrow n$ \KwTo $1$}{
        Generate a random integer $r$ between $1$ and $i$ (inclusive)\;
        \If{switching does not create self-to-self flows}{
            Switch the destinations of $f_r$ and $f_i$\;
        }
    }
}
\end{algorithm}
\vspace{1em}

For a candidate cluster in the original data, the \(p\)-value is estimated as \(R/(L+1)\), where \(R\) is the rank of its \(\mathit{LGLR}\) among the \(L\) maximum statistics obtained from the permutations. For example, if the cluster ranks 5th among 999 permutation maxima, the estimated \(p\)-value is \(0.005\).

\begin{figure}[htbp]
    \centering
    \includegraphics[width=0.8\textwidth]{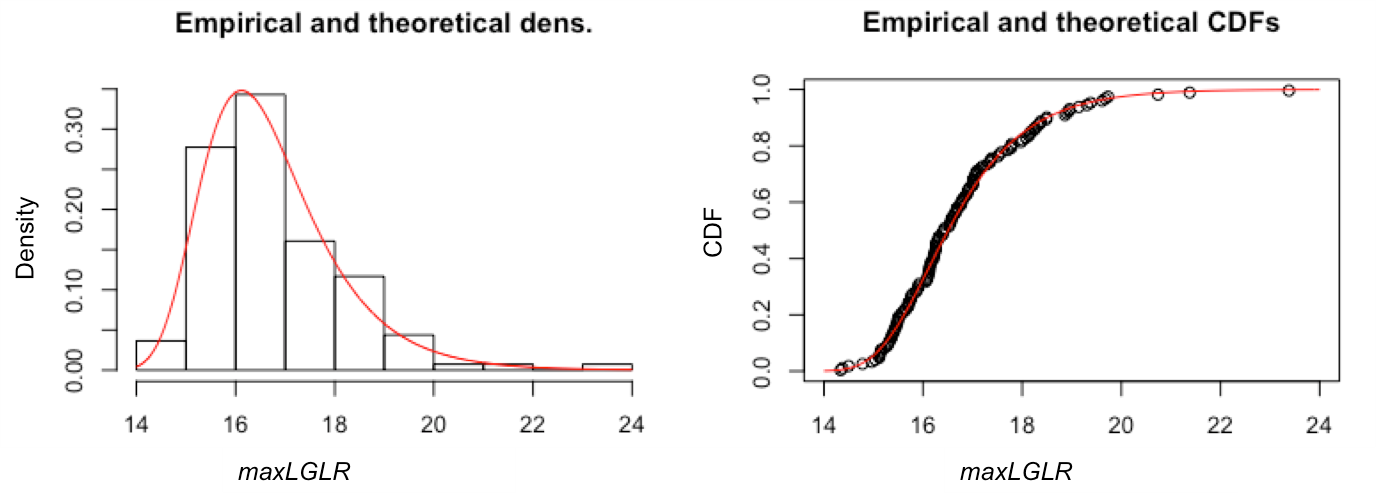}
    \caption{The empirical distribution of \(\mathit{maxLGLR}\) with 100 permutations of the senior migration data (\cref{sec:migration}) and the fitted Gumbel curves in red (location \(\mu = 16.12\) and scale \(\beta = 1.06\)).}
    \label{fig:empiricaldistribution}
\end{figure}

Because Monte Carlo simulation requires repeatedly running the scanning procedure, it can be computationally expensive. Empirically we observe that the distribution of the maximum \(\mathit{LGLR}\) values closely follows a Gumbel distribution \citep{gumbelStatisticsExtremes1958}, which models the distribution of maxima of random samples. The probability density function of the Gumbel distribution is 
\[
f(x)=\frac{1}{\beta}e^{-(z+e^{-z})}, \quad z=\frac{x-\mu}{\beta},
\]
where \(\mu\) and \(\beta\) are the location and scale parameters. As shown in \cref{fig:empiricaldistribution}, the empirical distribution of maximum \(\mathit{LGLR}\) values obtained from permutations closely matches the fitted Gumbel distribution. This allows accurate estimation of \(p\)-values using far fewer permutations (e.g., 50 rather than hundreds or thousands).

In practice, statistical testing is optional for the purpose of cartographic generalization. Because clusters are ranked by \(\mathit{LGLR}\), the top clusters selected for mapping typically represent only a small subset of all statistically significant clusters. Significance testing may therefore be applied when formal inference is required, while the ranking-based selection procedure alone is sufficient for producing flow maps.

\subsection{Cartographic Design for Cross-Scale Flow Mapping} 

The clusters identified by the selection procedure (\cref{alg:selection}) represent the dominant flow structures in the dataset. To visualize these clusters effectively, a cartographic representation is required that simultaneously conveys several attributes of each cluster. Each flow cluster contains six pieces of information: origin location, destination location, spatial scale of the origin region, spatial scale of the destination region, flow strength (measured by \(\mathit{LGLR}\) or its derived significance), and flow direction. 

Flow map design has long been a challenging problem \citep{toblerModelGeographicalMovement1981,guoFlowMappingMultivariate2009,guoOriginDestinationFlowData2014,koylu2017design,jenny2018design}. Conventional flow maps typically represent only origin, destination, and flow magnitude. In contrast, the clusters discovered by the cross-scale search involve two spatial scales, one at the origin and one at the destination, in addition to the statistical strength of the flow pattern. Therefore, a new symbol design is needed to encode these additional attributes while maintaining visual clarity.

The design of the cross-scale flow symbol follows several cartographic principles. First, the symbol should preserve the intuitive perception of movement between locations, which is commonly achieved through directional curves and arrows in flow mapping \citep{feketeOverlayingGraphLinks2003,koylu2017design,jenny2018design}. Second, it should visually encode multiple attributes without introducing excessive visual clutter. Third, the representation should remain compact so that multiple flow clusters can be displayed in a single map. Finally, the symbol should maintain geometric consistency with the spatial neighborhoods identified by the scan algorithm.

\begin{figure}[htbp]
    \centering
    \includegraphics[width=0.9\textwidth]{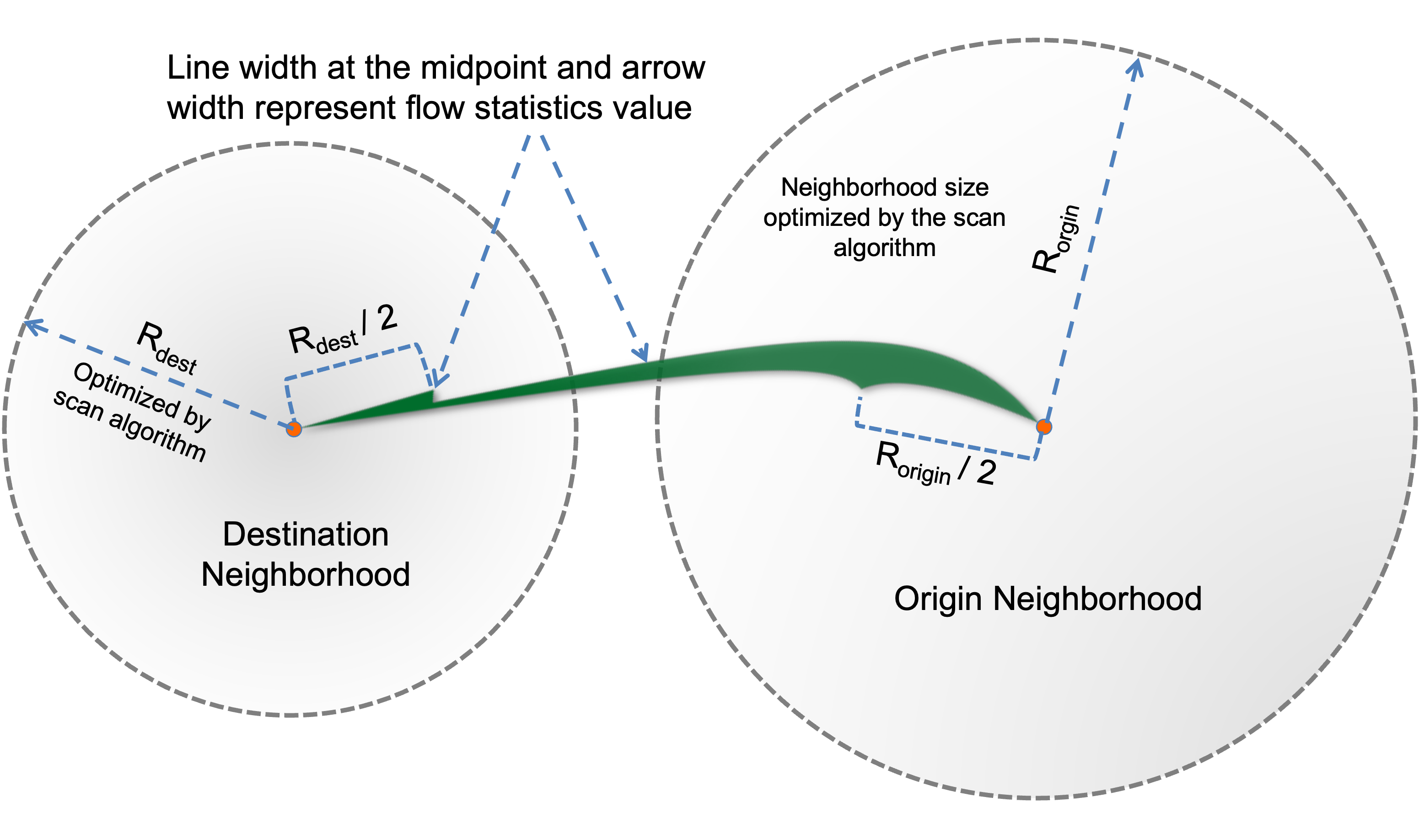}
    \caption{Flow symbol design for representing a cross-scale flow cluster.}
    \label{fig:symboldesign}
\end{figure}

\cref{fig:symboldesign} illustrates the flow symbol designed for representing a cross-scale flow cluster. The symbol integrates several geometric and visual elements to encode the six attributes of each cluster:

\begin{itemize}

\item \emph{Flow direction}.  
Flow direction is conveyed through multiple visual cues. The flow path is represented by a smooth Bézier curve connecting the origin and destination locations. The line gradually tapers from the origin toward the destination and terminates with an arrowhead, reinforcing the perceived direction of movement. Similar directional cues have been used in flow visualization \citep{holtenExtendedEvaluationReadability2011}. \textit{XFlowMap} can also support alternative flow-line symbols such as straight lines with arrowheads or animated flow movements; however, these alternatives cannot fully encode all cross-scale flow information represented in the proposed design.

\item \emph{Origin and destination locations}.  
The flow symbol begins at the centroid of the origin neighborhood and ends at the centroid of the destination neighborhood, maintaining a clear spatial reference to the underlying geographic locations.

\item \emph{Origin scale}.  
The spatial scale of the origin neighborhood identified by the scan algorithm is represented by the curvature segment extending from the origin toward the midpoint of the flow line. This segment visually reflects the radius of the origin neighborhood.

\item \emph{Destination scale}.  
The spatial scale of the destination neighborhood is represented by the length of the arrow segment at the destination, which is proportional to the radius of the destination neighborhood.

\item \emph{Flow strength}.  
The strength of the flow cluster, measured by \(\mathit{LGLR}\), is encoded using both line width and color intensity. The line thickness at the midpoint of the flow symbol and the width of the arrowhead increase with larger \(\mathit{LGLR}\) values. In addition, a sequential color scheme is used in which the intensity of a chosen hue increases across ordered class ranges of the strength measure.
\end{itemize}

These encodings employ multiple visual variables—including position, size, shape, orientation, and color—to represent the multidimensional attributes of cross-scale flow clusters. Together, they allow multiple attributes of a flow cluster to be interpreted simultaneously while maintaining a compact representation suitable for map display (\cref{fig:symboldesign}). The circles representing the origin and destination neighborhoods are shown in \cref{fig:symboldesign} for illustration purposes to clarify the geometric relationship between the flow symbol and the detected clusters. In the final flow map, these circles can be optionally displayed or hidden depending on the desired level of visual emphasis (see \cref{fig:seniorMigrationMap} and \cref{fig:SelectedSeniorClusters}).

In addition to static mapping, \textit{XFlowMap} naturally supports \textit{interactive exploration} of flow patterns across spatial scales. Because flow clusters are detected at multiple scales during the cross-scale search, the map can dynamically adjust the level of detail when the user zooms in or out. When the map is zoomed out, only large-scale clusters with stronger statistical support are displayed. As the user zooms into a region, additional local clusters can be revealed by automatically adjusting parameters such as $\mathit{maxSize}$ and the number of clusters selected for display. This enables scale-appropriate flow generalization without requiring manual reconfiguration of the dataset. Examples of such interactions are illustrated in \cref{fig:zoominmap}, which shows a zoomed-in view of the northeastern United States with additional clusters revealed at finer spatial scales.

Interactive exploration can also be supported for individual flows. When a specific flow is selected, the system can reveal its associated neighborhoods, including the origin and destination cycles detected during the scan process. Clusters at different scales surrounding the same focal flow or its neighboring flows can also be displayed. 

\section{Evaluation of XFlowMap Using Synthetic Data}

\begin{figure}[!htbp]
    \BlankLine
    \centering
    \includegraphics[width=1\textwidth]{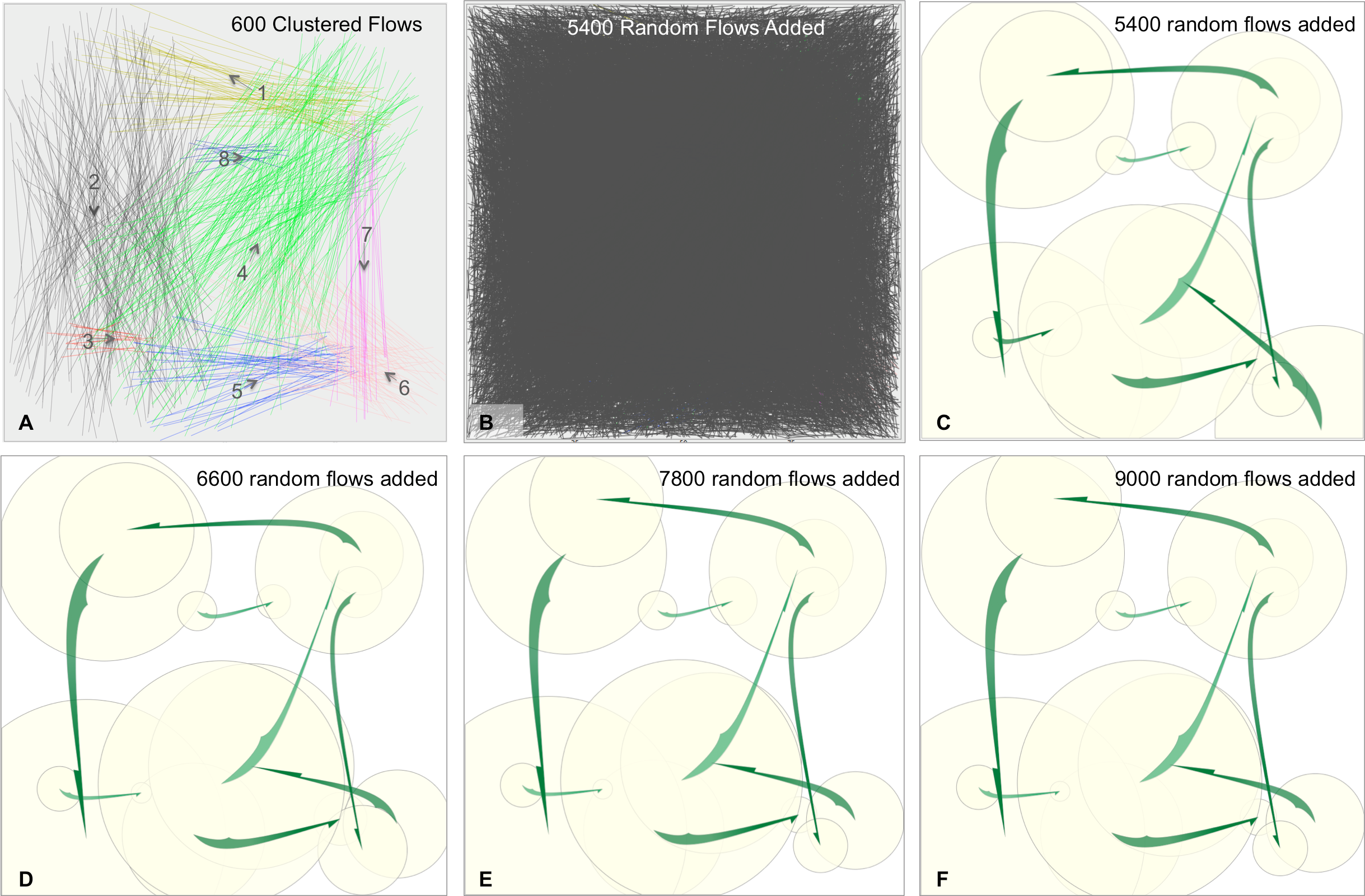}
    \caption{Experiments with synthetic datasets. (A) 600 flows in eight clusters; (B) 5400 random flows added; (C-F) significant clusters discovered for four synthetic datasets (with 5400, 6600, 7800, and 9000 random flows, respectively).}
    \label{fig:experimentsWithSyntheticData}
\end{figure}

To evaluate the proposed approach for detecting cross-scale flow patterns, we generate a series of synthetic datasets that contain known flow clusters at various scales together with a large number of random flows. Specifically, we first generate eight flow clusters across different scales, each consisting of flows from a circular origin area to a circular destination area of varying sizes, as shown in \cref{fig:experimentsWithSyntheticData}A:
\begin{multicols}{3}
\begin{itemize}
\item Cluster 1: 47 flows
\item Cluster 2: 193 flows
\item Cluster 3: 12 flows
\item Cluster 4: 193 flows
\item Cluster 5: 41 flows
\item Cluster 6: 80 flows
\item Cluster 7: 22 flows
\item Cluster 8: 12 flows
\end{itemize}
\end{multicols}

Then we add a large amount of random flows (\cref{fig:experimentsWithSyntheticData}B). To simulate random patterns at different spatial scales, we ensure that equal numbers of random flows are generated for different distance ranges. Four synthetic datasets are generated with progressively increasing levels of random flows (noise):
\begin{itemize}
\item Dataset 1: 600 clustered flows and 5400 random flows (90\%noise)
\item Dataset 2: 600 clustered flows and 6600 random flows (92\%noise)
\item Dataset 3: 600 clustered flows and 7800 random flows (93\%noise)
\item Dataset 4: 600 clustered flows and 9000 random flows (94\%noise)
\end{itemize}

We run our approach to scan, test, and select non-overlapping significant clusters from each dataset. For Dataset 1, eight significant clusters (\(p\text{-value} < 0.01\)) are detected, which correspond very well to the eight true clusters (\cref{fig:experimentsWithSyntheticData}C). The fitted Gumbel distribution for 100 permutations is: location \(\mu = 14.145\) and scale \(\beta = 0.763\). The cut value for \(p\text{-value} < 0.01\) is \(\mathit{LGLR} = 17.65\), which separates the eight discovered clusters from others of lower \(\mathit{LGLR}\) values, among which the highest is 14.01 (\(p\text{-value} > 0.5\)).

\cref{tab1} shows the significant LGLR values of the eight flow clusters and \cref{fig:experimentsWithSyntheticData}C-F present the maps of the eight discovered clusters, for each synthetic dataset. From Dataset 1 to Dataset 4, each dataset includes 1200 additional random flows. The results show that our approach can reliably discover the true clusters even with the presence of heavy noise, regardless of their size and scale. This indicates that the statistical measure is a reliable and robust indicator of flow cluster strength across scales. The scan algorithm together with the significance testing procedure effectively distinguishes true flow patterns from random noise. More importantly, the approach not only detects the locations of the clusters but also identifies their corresponding scales at the origin and destination separately. The minor differences among the four maps (\cref{fig:experimentsWithSyntheticData}C-F) are caused by the randomly added flows, which may slightly alter some clusters. Adding more noise generally weakens all clusters, particularly for the stronger ones (\cref{tab1}). For certain weak clusters, however, the \(\mathit{LGLR}\) value may improve a bit after adding more noise. This occurs because our method searches for clusters surrounding each existing flow, which may not exactly be at the cluster center. More random flows may help better position and detect the cluster.

\begin{table}
\tbl{Evaluation results for the four synthetic datasets. Signficant \(\mathit{LGLR}\) values are marked with \(^*\) (\(p\text{-value} < 0.01\)).}
{\begin{tabular}{llllllllll}
  \toprule
  Data &Cluster1 &Cluster2&Cluster3&Cluster4&Cluster5&Cluster6&Cluster7&Cluster8&Cluster9\\
  \midrule
  Dataset1 & 62.60\(^*\)& 55.64\(^*\)& 46.98\(^*\)& 33.10\(^*\)& 30.01\(^*\)& 23.04\(^*\)& 22.10\(^*\)& 18.13\(^*\)& 14.01 \\
  Dataset2 & 54.92\(^*\)& 54.91\(^*\)& 45.08\(^*\)& 33.02\(^*\)& 29.78\(^*\)& 23.71\(^*\)& 21.05\(^*\)& 20.89\(^*\)& 14.38 \\
  Dataset3 & 52.49\(^*\)& 49.21\(^*\)& 35.75\(^*\)& 26.75\(^*\)& 26.52\(^*\)& 26.32\(^*\)& 23.97\(^*\)& 18.87\(^*\)& 14.08 \\
  Dataset4 & 50.92\(^*\)& 44.76\(^*\)& 36.78\(^*\)& 26.39\(^*\)& 24.97\(^*\)& 24.15\(^*\)& 22.75\(^*\)& 18.05\(^*\)& 14.14 \\
  \bottomrule
\end{tabular}}
\label{tab1}
\end{table}

\section{Mapping U.S. Migration Flows with XFlowMap}
\label{sec:migration}

We apply the proposed flow cluster detection approach to U.S. county-level migration data to demonstrate its ability to reveal significant migration patterns across multiple spatial scales. The dataset is derived from the 2000 U.S. Census and records county-to-county migrants who moved during the five-year period 1995-2000. Although more recent migration data are available from the American Community Survey (ACS), those datasets provide only aggregate migration flows without demographic breakdowns. The 2000 Census dataset remains the most detailed publicly available migration flow dataset with county-level origin-destination information stratified by demographic characteristics, making it particularly suitable for methodological studies of flow pattern detection and visualization.

We focus on the migration within the contiguous United States, covering 3,075 counties across the 48 contiguous states and the District of Columbia. It is well understood that populations in different age groups tend to migrate differently. For example, many young people around age 18 move to attend college; people aged 25-29 often move for job opportunities; and senior populations tend to have different destination preferences. In this case study, we focus on two age groups: 65-69 and 25-29.

The migration data for the senior population (age 65-69) contains 916,856 migrants, with 70131 non-empty county-to-county flows among 3075 U.S. counties. Centered on each of the 70131 flows, the scan algorithm (\cref{alg:scan}) evaluates a total of 2,108,296,571 flow cluster candidates by combining different scales at the origin and the destination, with \(\mathit{maxSize} = 916856 / 5\) (meaning that the maximum neighborhood size evaluated for each location includes at most one-fifth of the total migrants). For each focal flow, the strongest flow cluster is recorded. The largest \(\mathit{LGLR}\) is 16326. To test the significance of these clusters, we run 100 permutations, obtain 100 maximum \(\mathit{LGLR}\) values, and fit a Gumbel distribution (location \(\mu = 16.12\) and scale \(\beta = 1.06\); see \cref{fig:empiricaldistribution}), based on which the threshold \(\mathit{LGLR}\) value is 20.1 for \(p\text{-value} = 0.01\) and 28.3 for \(p\text{-value} = 0.00001\).

\begin{figure}[htbp]
    \centering
    \includegraphics[width=1\textwidth]{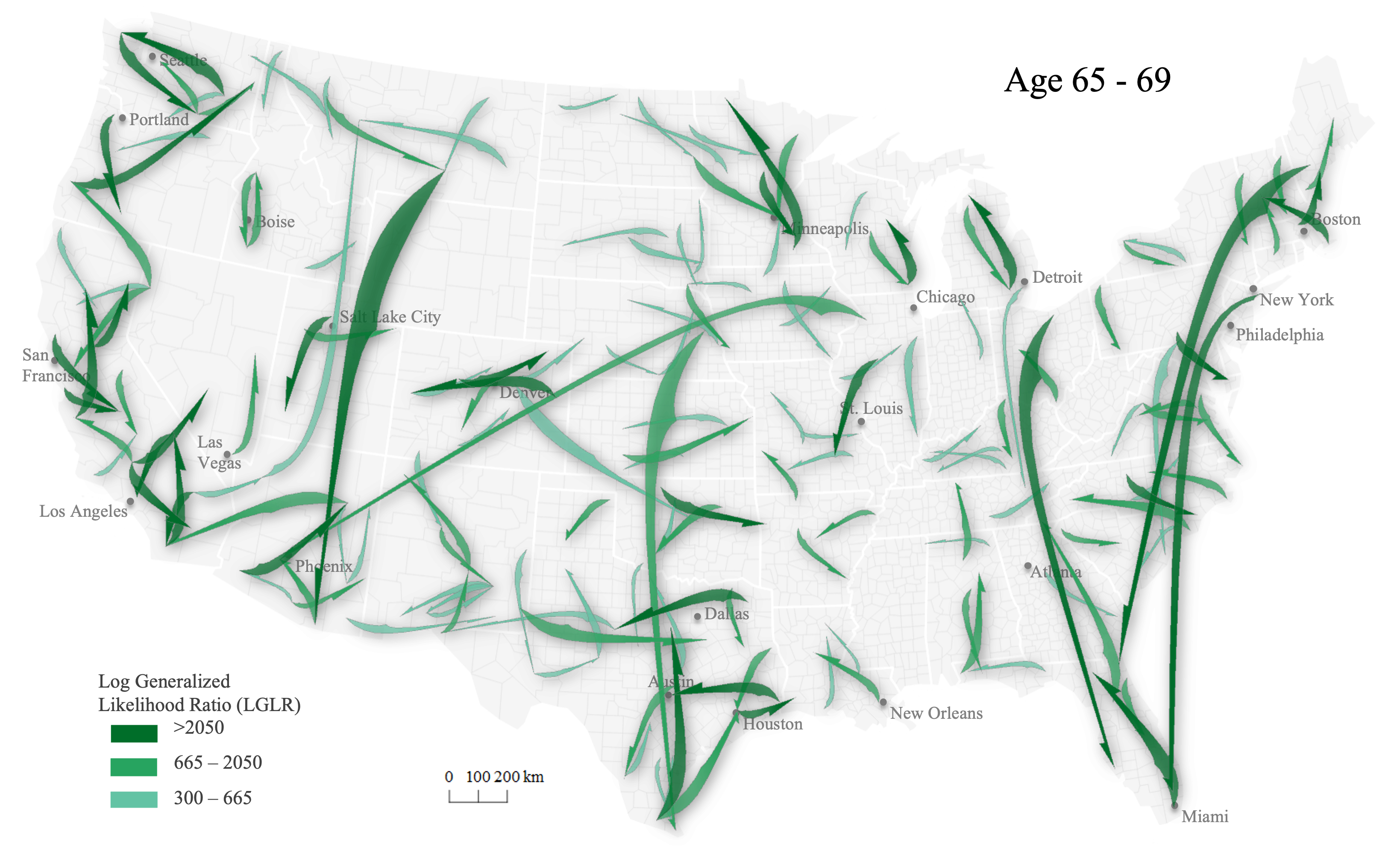}
    \caption{Non-overlapping significant flow clusters of senior migrants (age 65-69) with \(\mathit{LGLR}\) \textgreater{} 300 and distance \textgreater{} 200km.}
    \label{fig:seniorMigrationMap}
\end{figure}

After removing overlapping clusters (\cref{alg:selection}), there are 7942 significant clusters with \(p\text{-value} < 0.01\), among which about half are short-distance (\textless{} 200km) local flow clusters. One can choose to map the most important clusters if there are too many to visualize in a single map. One may also focus on long-distance clusters if the focus is on national trends. For example, \cref{fig:seniorMigrationMap} shows 166 flow clusters of senior migrants, with \(\mathit{LGLR} > 300\) and distance \textgreater{} 200km. These two thresholds (i.e., \(\mathit{LGLR}\) and distance) are purely for mapping purposes and do not affect the major patterns---they control how much detail (i.e., weaker and shorter clusters) is included in the flow map.

\begin{figure}[!htbp]
    \centering
    \includegraphics[width=1\textwidth]{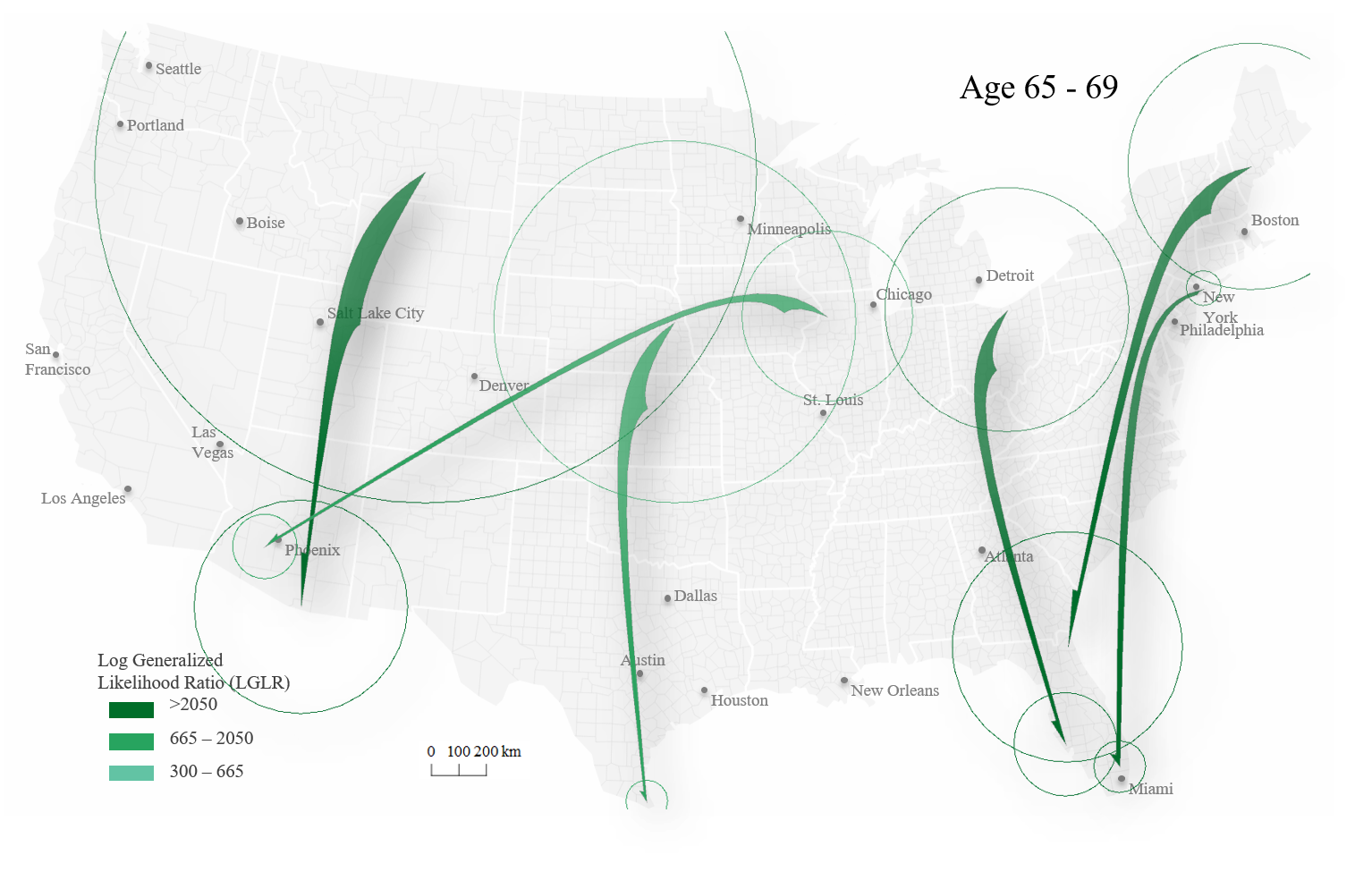}
    \caption{Six selected clusters from \cref{fig:seniorMigrationMap}, with origin/destination circles shown, to help interpret the mapped clusters.}
    \label{fig:SelectedSeniorClusters}
\end{figure}

From the flow map of clusters in \cref{fig:seniorMigrationMap}, we can obtain a holistic and yet statistically accurate understanding of the national trends and patterns of senior migration. To help illustrate how to interpret the mapped clusters, \cref{fig:SelectedSeniorClusters} shows six selected clusters from \cref{fig:seniorMigrationMap} with origin/destination circles shown. It is obvious that there is a strong migration trend from the northwestern states (Washington, Montana, Wyoming, Idaho, Utah, and others) to Southern Arizona. There are also a few strong in-migration flow clusters to Florida. Particularly, the New England area has a general preference of Florida and Southern Georgia/South Carolina; migrants from New York City have a strong focus on Miami; and senior migrants from the midwestern states such as Ohio and Michigan favor the Tampa Bay area. It is also interesting to see northwards trends from the South Florida to the northern part and from Florida/Southern Georgia to the Smokey Mountains. In the surrounding area of Washington, D.C., there is a clear migration trend to the coast; in California, migration generally occurs from coastal areas to inland regions; in Texas, migration trends generally point toward the central part of the state. These are just a few examples of the broader trends that can be observed from the single map in \cref{fig:seniorMigrationMap}.

\begin{figure}[htbp]
    \centering
    \includegraphics[width=0.9\textwidth]{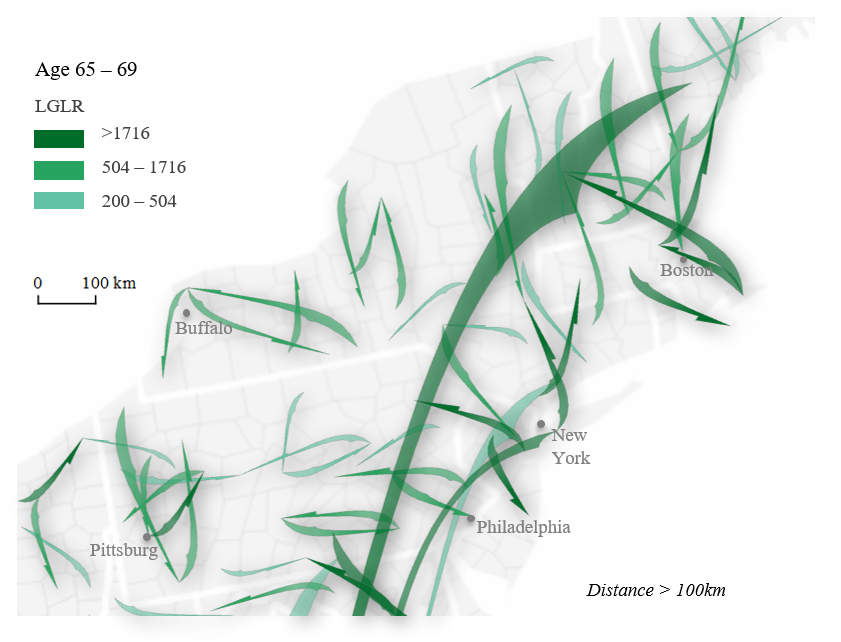}
    \caption{A zoomed-in flow map with additional details (\(\mathit{LGLR} > 200\) and distance \textgreater{} 100 km).}
    \label{fig:zoominmap}
\end{figure}

Our approach not only reveals patterns at different scales simultaneously but also supports interactive mapping with dynamic scale-based generalization. When the user zooms in on a specific region (e.g., New England), based on the map extent and scale, additional details can be revealed by lowering the \(\mathit{LGLR}\) and distance thresholds (i.e., adding more relatively weak and short flow clusters). For example, \cref{fig:zoominmap} shows a zoomed-in map of the Northeast with \(\mathit{LGLR} > 200\) and distance \textgreater{} 100km, which keeps all the flow clusters in \cref{fig:seniorMigrationMap} and adds more details locally. We can see strong outward migration trends from NYC and an interesting trend to the center region of New England from both the coast and the north border.

\begin{figure}[!htbp]
    \centering
    \includegraphics[width=1\textwidth]{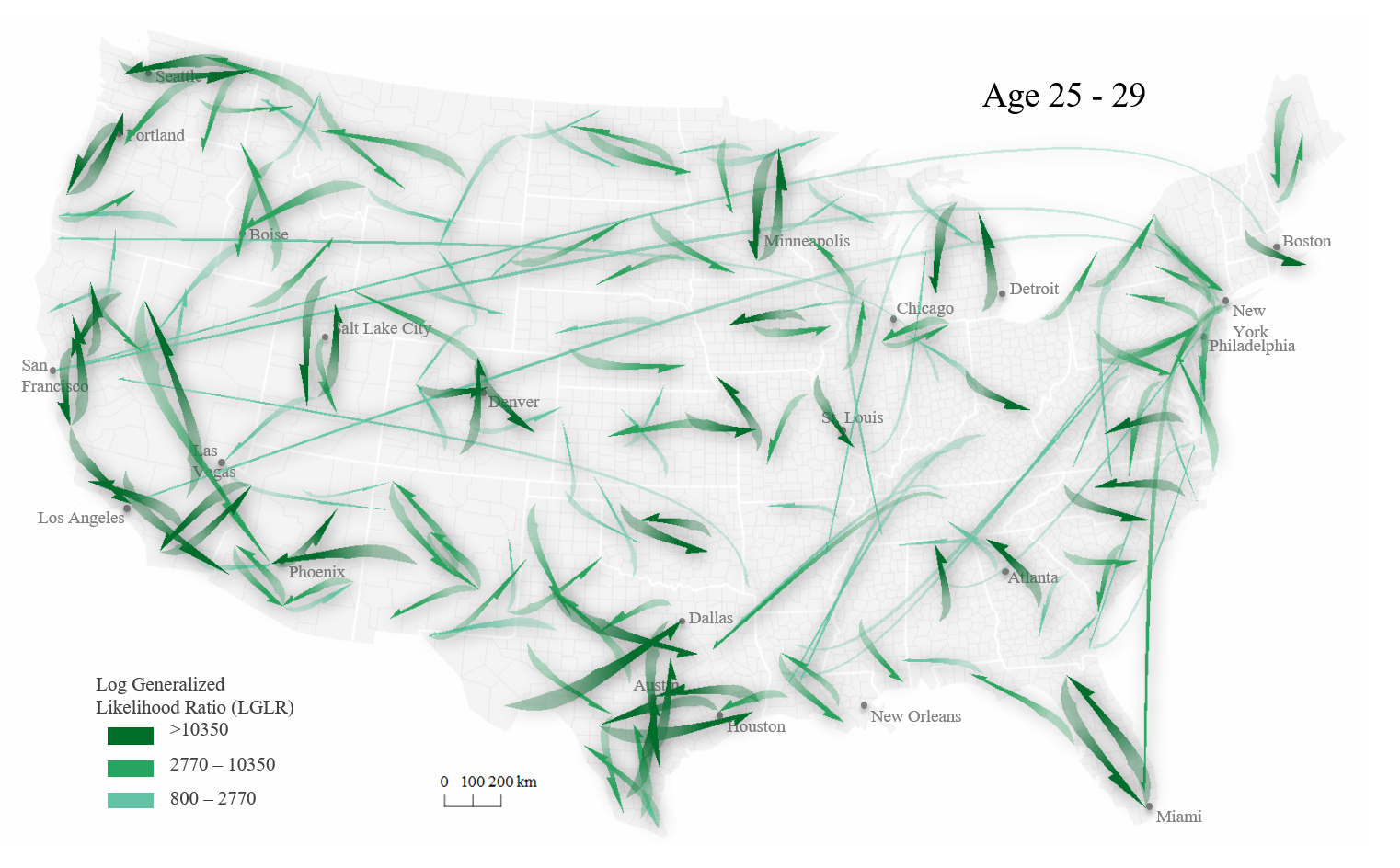}
    \caption{Flow clusters of young migrants (age 25-29) with \(\mathit{LGLR} > 800\) and distance \textgreater{} 250km. Unlike the map in \cref{fig:seniorMigrationMap,fig:SelectedSeniorClusters}, the origin half of each flow is rendered semi-transparent.}
    \label{fig:youngmigrationmap}
\end{figure}

\cref{fig:youngmigrationmap} shows the flow clusters of young migrants (age 25-29). Obviously, young people migrate in very different ways. Most strong clusters occur at short distances, meaning that young migrants primarily move locally, e.g., within states. While senior migrants show a dominant north-to-south migration pattern, long-distance migration among young people tends to follow a more east-west pattern, e.g., from Boston and New York to the San Francisco area and Los Angeles. With these two case studies, we hope to demonstrate the flexibility and robustness of our approach in extracting flow patterns at their most appropriate scales and rendering a holistic view of complex flow patterns in large spatial flow data.

\section{Conclusion and Discussion}

This paper presents \textit{XFlowMap}, a framework for the cross-scale generalization and mapping of large origin-destination flow datasets. Instead of mapping all individual flows directly, \textit{XFlowMap} extracts salient flow patterns, identifies their appropriate origin and destination scales, and renders them in a new flow map representation that supports a holistic interpretation of complex origin-destination flow data. The method works with both area-based and point-based flow data.

The proposed approach offers several advantages for flow mapping and generalization. First, it identifies flow patterns across multiple spatial scales rather than relying on a fixed aggregation level, thereby reducing limitations imposed by predefined spatial units. Second, it evaluates candidate flow patterns using a statistically guided measure so that flows associated with regions of different sizes can be compared in a consistent manner. Third, the resulting flow map provides a compact and information-rich synthesis of the original dataset, simultaneously encoding origin, destination, flow direction, flow strength, and the spatial scales of the two regions. The synthetic and migration case studies demonstrate that the method can effectively reveal meaningful patterns in large and noisy flow datasets.

Compared with the flow smoothing and density-based approach of \citet{guoOriginDestinationFlowData2014}, which detects patterns at a chosen scale determined by a bandwidth parameter, \textit{XFlowMap} systematically searches across origin and destination scales and can therefore detect patterns both at and across different spatial extents. The method also performs well for sparse and noisy flow data and rare events, making it possible to stratify datasets by attributes and compare flow patterns across different groups, as illustrated in the migration case study. In addition, the framework naturally supports interactive exploration, allowing the map to reveal flow patterns at appropriate levels of detail as the user zooms in or out.

Further research is needed on both cross-scale flow pattern identification and cartographic visualization, as the automated analysis and mapping of massive flow data across scales remains a new frontier open for innovation. The current framework may be extended in several directions. Different null models may be incorporated for different analytical purposes, and the permutation strategy may be refined to better reflect realistic movement processes such as group migration. In addition, although circular neighborhoods provide a practical balance between detection power and computational efficiency, other neighborhood shapes may also be explored in future work. Nevertheless, circular neighborhoods are advantageous for interpretation and cartographic design because they are easy to understand and require only a single parameter (the radius) to define. Future work should also evaluate the usability and effectiveness of the proposed flow map design through controlled user studies and task-based experiments. Such evaluations would help assess how well analysts and map readers can interpret cross-scale flow patterns and interact with the visualization. Overall, \textit{XFlowMap} provides a new framework for the automated discovery, generalization, and visualization of cross-scale flow patterns in large origin-destination datasets.

\section*{Data Availability Statement}

The data are available in the Figshare repository. During peer review, the dataset can be accessed via the following private link: \url{https://figshare.com/s/e96833b9f078b6104f30}. A reserved DOI has been assigned: \url{https://doi.org/10.6084/m9.figshare.31702921}. The dataset will be made publicly available upon acceptance of the manuscript.

\section*{Funding}

This work was supported in part by the National Science Foundation under Grant No. 0748813.  

\bibliographystyle{apacite}
\bibliography{cross-scale_flow_mapping_01}

\end{document}